# Optimization of Argon Plasma Working Pressure through Parallel PIC Simulations for Enhancement of Material Surface Treatment

Hadi Barati[1], Ali Torkaman[1], and Mehdi Fardmanesh[1]

[1]Department of Electrical Engineering, Sharif University of Technology, Tehran, Iran

Corresponding author:
Hadi Barati
Department of Electrical Engineering, Sharif University of Technology, Tehran, Iran
Email: baratihadi6@gmail.com , hadi.barati@ee.sharif.edu

**Abstract**

In this study, a novel method for simulating plasma dynamics using parallel programming has been developed. The equations based on Particle-in-Cell (PIC) method were utilized and adapted for this purpose. We utilized 35 processors from Sharif High Performance Computing (HPC) center and divided the plasma volume into 35 parts, with each part's PIC equation solved on a separate processor. Once the computations were completed, the results from all processors were combined to form a complete plasma volume. The simulations revealed that there is an optimal pressure for argon, at which the ion flux onto the electrode surface is maximized. Increasing the absolute value of the electrode potential also increases this flux. Therefore, for a given potential, selecting the optimal pressure is crucial for the most effective surface modification using argon plasma. In this work, for applied voltage of -500 V, the optimum pressure was 100 mTorr.

**1. Introduction**

Long-term success of medical implants depends on the integration of implant materials with biological tissues [1, 2]. In this process, surface characteristics of implant materials are quite important since they affect cellular reactions including adhesion, proliferation, and differentiation [2, 3]. Different surface modification methods have been investigated to improve the interaction between biological systems and implant surfaces; plasma surface treatment has come out as a possible solution [4, 5]. This approach improves biocompatibility by allowing the modification of the implant surface without changing its bulk characteristics [5]. The application of argon plasma treatment has been verified as a great enhancing method for both biological and surface properties of metallic and polymeric biomaterials. Hence, several areas of biocompatibility can be affected by argon plasma treatment.

Studies have shown that the argon plasma treatment processes, could enhance the hydrophilicity and wetting of implant materials, specifically for polyimide (PI) [6] and titanium [3]. For example, the surface free energy of PI surfaces nearly doubled with plasma treatment [6]. For collagen-based and poly (lactic-co-glycolic acid) (PLGA) scaffolds, argon plasma treatment decreases hydrophobicity, alters surface chemistry, and allows for partial collagen crosslinking, thereby resulting in diminished protein release and increased scaffold stability—in parameters that greatly control bone tissue engineering [7]. Similarly, in situ argon plasma treatment on 3D-printed polylactic acid (PLA) scaffolds has been documented as enhancing surface roughness and adding oxygen-containing functional groups but, in turn, diminishing the water contact angle and enhancing adhesion, proliferation, and viability of human adipose-derived stem cells [8]. In metallic substrates such as nanostructured titanium, argon plasma has been found effective in eliminating carbon contaminants, increasing hydrophilicity, and enhancing protein adsorption and osteoblast adhesion without influencing nanoscale morphology, thereby improving enhanced osseointegration [9]. Overall, these results support the inference that argon plasma at the same time manages surface chemistry and topography but, as such, develops a favorable microenvironment that promotes protein adsorption, cell adhesion, and osteogenic differentiation and thereby boosts the biocompatibility of various different implant materials. Besides chemical functionalization, plasma treatment can also modify the physical attributes of the surface of implants, i.e., geometry and roughness, which in turn regulate cellular activity [4, 10]. Variation of nanoscale topography can alter the adsorption kinetics of proteins and impose cellular orientation, spreading, and differentiation directly [3, 6]. Therefore, quantitative insights into the mechanisms by which plasma-induced changes in morphology contribute to biocompatibility require simulation-based analysis, such as the Particle-in-Cell (PIC) technique, for investigating plasma–surface interactions at the microscopic scale [8, 7].

Although several studies have investigated the effects of argon plasma on biocompatibility, the optimization of its treatment parameters still remains an area of continuous research. An understanding of the interplay between plasma properties such as ion energy, density, and surface interaction and their implications on functionalization can significantly facilitate implant surface property enhancement as well as biological response improvement. In this regard, PIC simulation utilization could provide valuable insight into plasma creation, argon ion and radical kinetics as well as their material surface interaction behavior, thus contributing towards the rationalizing plasma process development towards biocompatibility optimization. Thus, the development of implant surfaces with enhanced biological performance could be greatly advanced by research concentrating on the optimization of argon plasma through PIC simulations.

## 2. PIC description

In the current research, the PIC method is employed for simulating the plasma particle motions. With this method, the time-evolved population of the plasma ions and electrons hitting the substrate surface can be determined. This value is a key variable in assessing the effectiveness of the surface bombardment by the plasma for the purpose of surface functionalization.

In PIC, the plasma is modeled as a mixture of computational (super) particles, including ion and electron super-particles. A super-particle represents a large number ($N_p$) of real plasma particles with similar energetic and spatial properties. These super-particles are typically treated as having a finite spatial extent or shape. This finite size is crucial as it models the collective behavior of the $N_p$ real particles, reduces unphysical short-range interactions between individual super-particles, and allows the system to mimic a weakly coupled plasma with a computationally tractable number of super-particles.

Here, a non-collisional partially ionized plasma is considered [], and thus the super-particle distribution throughout the plasma domain can be determined by solving the governing equation of Maxwell-Vlasov:

$$\frac{\partial F_p}{\partial t} + \vec{v} \cdot \nabla_{\vec{R}} F_p + q_p(\vec{E} + \vec{v} \times \vec{B}) \cdot \nabla_{\vec{p}} F_p = 0 \qquad (1)$$

$F_p(\vec{R}, \vec{V}, t)$ is the probability density distribution function in phase space for a super-particle (representing species p, such as electrons or ions) with charge $q_p$ and momentum $\vec{p} = m_p \vec{V}$. $\vec{V}$ is the velocity vector of each super-particle. $\vec{E}$ and $\vec{B}$ are the macroscopic electric and magnetic fields.

The distribution function for a single super-particle $k$ located at $\vec{r}_k(t)$ with velocity $\vec{V}_k(t)$ is often represented as

$$F_k(\vec{R}, \vec{V}, t) = N_p \varphi_x\left(\vec{R} - \vec{R}_k(t)\right) \varphi_v(\vec{V} - \vec{V}_k(t)) \qquad (2)$$

where $\varphi_x$ is a spatial shape function and $\varphi_v$ is a velocity shape function (typically a Dirac delta function, implying all $N_p$ constituent real particles share the velocity $\vec{V}_k$). The total distribution function for species $p$ is

$$F_p = \sum_k F_k \tag{3}$$

In the PIC approach, the Vlasov equation is not directly solved for $F_p$. Instead, the trajectories of a large number of super-particles are tracked using the characteristic equations (equations of motion), and these particles serve as sources for Maxwell's equations.

$$\frac{\partial \vec{R}_p}{\partial t} = \vec{V}_p \tag{4}$$

$$\frac{\partial \vec{V}_p}{\partial t} = \frac{q_p}{m_p}\left(\vec{E}_p + \vec{V}_p \times \vec{B}_p\right) \tag{5}$$

Here, $\vec{r}_p$ and $\vec{v}_p$ are the position and velocity of an individual super-particle. $\vec{E}_p$ and $\vec{B}_p$ are the electric and magnetic fields at the super-particle's position $\vec{R}_p$. These fields are not the direct microscopic fields but are interpolated from the macroscopic fields $\vec{E}_g$ and $\vec{B}_g$ calculated on a computational grid.

Maxwell's equations govern the evolution of these macroscopic fields:

$$\nabla_g \times \vec{E} = -\frac{\partial \vec{B}}{\partial t} \tag{6}$$

$$\nabla_g \times \vec{B} = \mu_0 \vec{J}_g + \frac{1}{c^2}\frac{\partial \vec{E}}{\partial t} \approx \mu_0 \vec{J}_g \tag{7}$$

$$\nabla_g \cdot \vec{E} = \frac{\rho_g}{\varepsilon_0} \tag{8}$$

$$\nabla_g \cdot \vec{B} = 0 \tag{9}$$

The approximation in Equation (7) implies that the displacement current term ($\frac{1}{c^2}\frac{\partial \vec{E}}{\partial t}$) is neglected, often valid for non-relativistic plasmas or low-frequency phenomena.

For the PIC simulation, the plasma region is discretized into a mesh of computational grid points. At these points, Equations (6)–(9) are solved. The subscript "g" on operators indicates differentiation performed at the grid points.

The source terms $\rho_g$ (charge density) and $\vec{J}_g$ (current density) are calculated by summing contributions from all super-particles onto the grid.

The contribution of each super-particle to the grid-based densities is determined using a weighting or shape function. This function depends on the relative distance between the grid point position $\vec{R}_g$ and the super-particle position $\vec{R}_p$. In this research, the spatial shape of a super-particle itself is conceptually based on the zero-order b-spline function (a rectangular or "top-hat" profile in each dimension):

$$\emptyset_0(\vec{R}_g - \vec{R}_p) = \prod_{i \in \{x,y,z\}} \begin{cases} 1 & if \frac{|\vec{R}_{g,i} - \vec{R}_{p,i}|}{\|\Delta \vec{R}_i\|} < 0.5 \\ 0 & O.W. \end{cases} \tag{10}$$

where $\Delta R_i$ is the grid cell size in dimension $i$. This $b_0$ function defines the "cloud" of a super-particle. More generally, for a 1D coordinate $\xi$, $\emptyset_0(\xi) = 1$ if $\frac{|\xi|}{\Delta \xi} < 0.5$ and 0 otherwise.

The current density $\vec{J}_g$ and charge density $\rho_g$ at a grid point $\vec{R}_g$ are calculated by summing the contributions of all $N$ super-particles, weighted by an interpolation function W:

$$\vec{J}_g(\vec{R}_g) = V_g^{-1} \sum_{k=1}^{N} q_k \vec{V}_k \Delta(\vec{R}_g - \vec{R}_k) \tag{11}$$

$$\rho_g(\vec{R}_g) = V_g^{-1} \sum_{k=1}^{N} q_k \Delta(\vec{R}_g - \vec{R}_k) \tag{12}$$

$V_g$ is the control volume (cell volume) associated with the grid point. $q_k$ (Same as $q_i$) and $\vec{V}_k$ (same as $\vec{V}_i$) are the charge and velocity of the k-th super-particle.

The interpolation function $\Delta(\vec{R}_g - \vec{R}_k)$ determines how a particle's charge and current are distributed to nearby grid points. It is related to the particle shape function. For a $\emptyset_0$ (0th-order b-spline) particle shape, the commonly used Cloud-In-Cell (CIC) method employs a weighting function W that is a 1st-order b-spline, $\emptyset_1$. This $b_1$ function can be obtained by convolving two $\emptyset_0$ functions:

$$\Delta(\vec{R}_g - \vec{R}_k) = \emptyset_1(\vec{R}_g - \vec{R}_k) = \int_{-\infty}^{\infty} d\vec{r'} \emptyset_0(\vec{R}_g - \vec{R}_k - \vec{r'})\emptyset_0(\vec{r'}) \tag{13}$$

In multiple dimensions, $\Delta$ is typically a product of 1D $b_1$ functions, e.g.,

$$\Delta(\vec{r}_g - \vec{r}_p) = \emptyset_1\left(\frac{x_g - x_p}{\Delta x}\right)\emptyset_1\left(\frac{y_g - y_p}{\Delta y}\right)\emptyset_1\left(\frac{z_g - z_p}{\Delta z}\right) \tag{14}$$

This weighting scheme ensures that $\sum_g \Delta(\vec{R}_g - \vec{R}_k) = 1$ for any $\vec{r}_p$, conserving charge.

The fields $\vec{E}_p$ and $\vec{B}_p$ acting on a super-particle are obtained by interpolating the grid-based fields $\vec{E}_g, \vec{B}_g$ to the particle's position $\vec{R}_p$.

In the present work, the bombardment of a material surface by plasma particles is investigated. Since the electrode shape of the experimental apparatus is disk-shaped (schematic in Figure 1), the cylindrical coordinate system $(r, \theta, z)$ is chosen for expanding and solving the governing equations. Hence, it is assumed that the plasma is generated inside a cylindrical vacuum chamber and over an DC-powered electrode. The initial pressure of working gas, i.e., argon is assumed as the working pressure of the plasma.

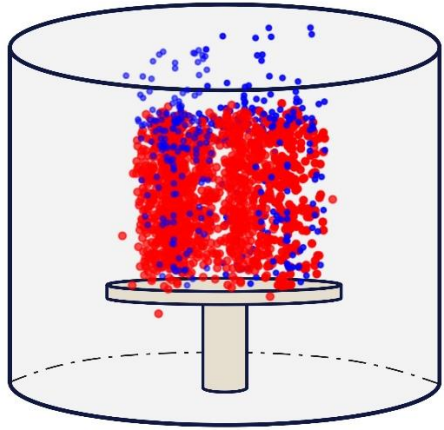

Figure1. Cylindrical vacuum chamber. The plasma is formed over the electrode. The plasma ions and electrons are shown in red and blue, respectively.

For the field solution step, particularly for the electric field which often dominates low-temperature plasma-surface interactions, and when boundary conditions are defined by electrode potentials, it is practical to solve Poisson's equation $\vec{\nabla}^2 \Phi = -\frac{\rho_g}{\varepsilon_0}$ to find the distribution of electric potential $\Phi$ throughout the plasma domain. (Laplace's equation $\vec{\nabla}^2 \Phi = 0$ is a special case for $\rho_g = 0$). In cylindrical coordinates:

$$\vec{\nabla}^2 \Phi = \frac{1}{r}\frac{\partial}{\partial r}\left(r\frac{\partial \Phi}{\partial r}\right) + \frac{1}{r^2}\frac{\partial^2}{\partial \theta^2}(\Phi) + \frac{\partial^2}{\partial z^2}(\Phi) = -\frac{\rho_g}{\varepsilon_0} \tag{15}$$

After solving for Φ on the grid, the electric field components are found by $\vec{E}_g = -\nabla_g \Phi$.

In PIC method, numerical stability and accuracy are paramount. This is to say selecting a time step δt satisfying the stability criteria (e.g., Courant-Friedrichs-Lewy condition, to resolve plasma/cyclotron frequencies), choosing sufficiently many super-particles per cell so that statistical noise is insignificant, and choosing accurate, conservative interpolation schemes. Boundary conditions for both particles and fields should be handled with utmost care. Traditionally, for simplicity, PIC methods consider super-particles of fixed shape, and strictly speaking, the phase-space volume of an individual super-particle element cannot be conserved as required by Liouville's theorem for the element itself (even though the system as a whole strives to satisfy it).

## 3. Multi-process (parallel) computation

For the simulations of the present work, we have employed the Sharif University HPC (High Performance Computing) system. Since, in an HPC approach, computation can be performed in parallel manner, the computation code should follow parallel multi-process programming method. In this purpose, we have developed a novel parallel programming method to compute the evolution of the plasma particles motions utilizing multiple cores or processes. As illustrated in Figure2, the plasma column is divided to n sectors where n is the number of cores or processors available by the HPC system. In the present work, 35 cores or processes, i.e., n=35, have been employed from HPC for the simulations while for the illustration purposes the plasma column has divided into 10 sectors in Figure 2. Since sectors are simulated on individual cores or processors, the marginal plasma particles cannot influence each other. To alleviate this problem, the sectors are marginally overlapped and thus, the marginal particles can exert force on each other and this can make the simulation results more consistent with the plasma dynamics. However, the marginal effects are considered twice and thus the average position and velocity of a marginal particle is considered.

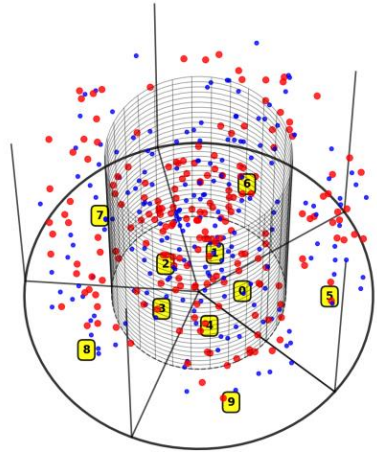

Figure 2. the plasma column is divided to n sectors, for example n=10. The motion evolution of plasma particles located in each sector is simulated on a different core (processor).

## 4. Results and discussion

The effects of the working pressure of the argon plasma and the electrode voltage have been investigated through multi-processing simulations. These effects on the ion beam flux impeding the electrode surface are shown in Figures 3 and 4, respectively. the amount of the pressure influence on the mass and charge of super-particles through the following equations [11]

$$m_{si} = \frac{V_{plasma}\rho_{Ar,0}}{N_i} P_{Ar} \tag{16}$$

$$m_{se} = \frac{m_e V_{plasma}\rho_{Ar,0}}{M_{Ar}N_i} P_{Ar} \tag{17}$$

$$q_{si} = \frac{Z_{eff}eV_{plasma}\rho_{Ar,0}}{M_{Ar}N_i} P_{Ar} \tag{18}$$

$$q_{se} = \frac{eV_{plasma}\rho_{Ar,0}}{M_{Ar}N_i} P_{Ar} \tag{19}$$

$m_{si}$ and $m_{se}$ are super-ion and -electron masses, respectively. $\rho_{Ar,0}$ is the argon mass density at 1 atmosphere pressure. $P_{Ar}$ is the filling pressure of argon. $V_{plasma}$ is the argon plasma volume. $M_{Ar}$ is the argon atomic mass. $N_i$ is the number of super-ions that is set manually. The maximum value of $N_i$ is restricted by the HPC computational limitations. Due to Sharif HPC available computational resources, the maximum value of $N_i$ was set equal to 10000. Thus, for partially ionized argon plasma with effective charge number $Z_{eff}$ equal to 11 [12], the number of super-electrons becomes 110000. For the simulations, the electrode potential has been set as various dc values. Thus, the super-ions are attracted by the electrode while the super-electrons are repelled.

As seen in figure 3, the plasma ions exhibit a maximum average hitting-electrode flux of 1.4309x10$^{15}$ (#/m$^2$ s$^{-1}$) at pressure of 100 mTorr with respect to pressures of 1, 10, 200 mTorr. Thus, this pressure can be seen as an optimum working pressure at the electrode potential of -500 V for using argon plasma for material surface functionalization. The average flux at pressures of 1,10, and 200 mTorr are as 1.2936 x10$^{15}$, 1.4293 x10$^{15}$, and 1.3210 x10$^{15}$ (#/m$^2$ s$^{-1}$), respectively.

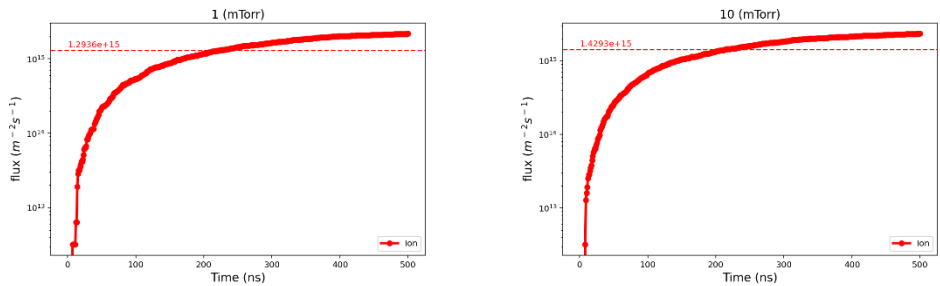

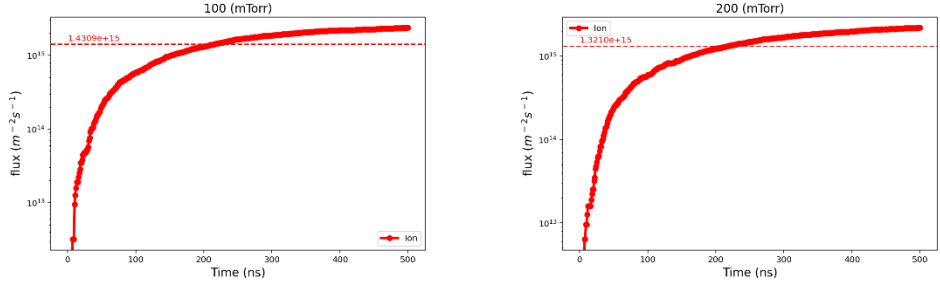

Figure 3. Effect of pressure on ion flux receiving on the electrode surface. The pressure value is shown in title of each case (V=-500 V).

For investigation the electrode potential on the ion flux bombarding the electrode surface, the working pressure is set as 100 mTorr and simulations are performed for various values of electrode potentials. The obtained results are illustrated in Figure 4. As seen, increasing the more negative potential, the more flux is obtained. The average flux is 6.327 $\times 10^{14}$, 1.1055 $\times 10^{15}$, 1.4309$\times 10^{15}$ (Figure3), 1.5132$\times 10^{15}$, and 1.6568$\times 10^{15}$ (#/m$^2$ s$^{-1}$) for -100, -300, -500, -700, and -1000 V, respectively.

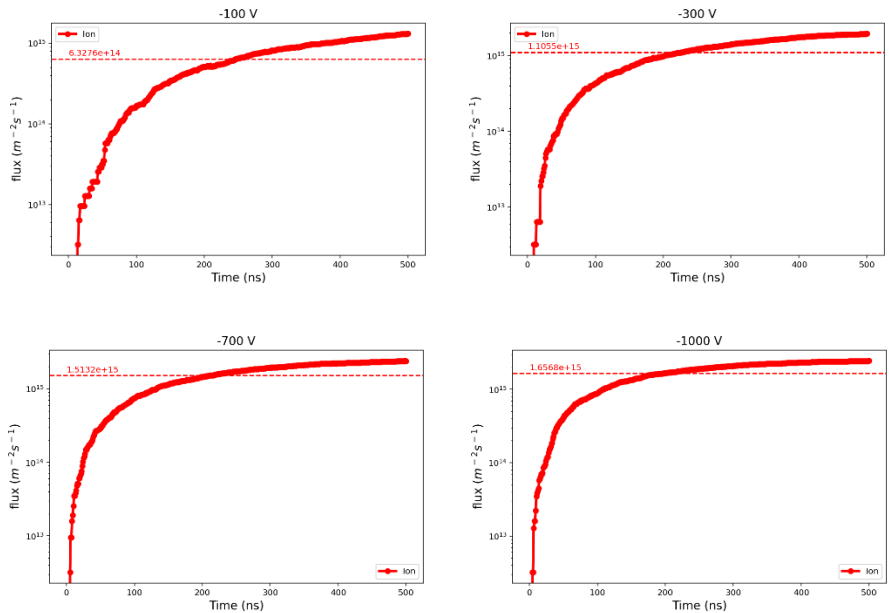

Figure 4. Effect of electrode voltage on ion-flux receiving on the electrode surface. The potential value is shown in title of each case (P=100 mTorr).

Increasing the amplitude of the electrode voltage enhances the intensity of ion bombardment on the electrode. However, there is an optimal pressure value for this process. In essence, a more negative electrode potential improves the modification of the material surface by the argon plasma. Nevertheless, technical limitations may restrict the maximum potential that can be applied to the electrode, so the pressure must be adjusted accordingly to optimize the available potential.

## 5. Conclusion

In this work, a novel approach for simulation plasma dynamics utilizing parallel programming method has been developed. The PIC based equations have been considered and developed for this purpose. We have employed 35 processors from Sharif HPC and divided the plasma volume to 35 parts that the PIC equation for each part has been numerically solved on a different processor. After the computations finished, the results of all cores have been assembled to one plasma volume. The simulations showed that for the argon pressure there is an optimal value at which the ion flux into the electrode surface maximizes while increasing the absolute of the electrode potential increases this flux. Thus for a set potential, the optimum pressure should be adopted for the most effective surface modification by the argon plasma.